# Plasmon drag effect in profile-modulated gold films. Theory and experiment

N. Noginova[1], M. LePain[2], V. Rono[1], S. Masshadi[1], R. Hussain[3], M. Durach[2]

[1]*Norfolk State University, Norfolk VA*
[2]*Georgia Southern University, Statesboro GA*
[3]*Intel Corporation, Hillsboro OR*

*Abstract*. Plasmon drag effect is studied theoretically and experimentally in gold films with a sine-wave height profile. Numerical simulations based on the modified electromagnetic momentum loss approach are shown to correctly describe the photoinduced voltage associated with propagating surface plasmon polaritons. Direct proportionality of energy and momentum transfer in interactions of plasmons and free electrons in metal is proven to be valid for surfaces with relatively low height modulation amplitudes. The effect is also discussed in frames of an equivalent circuit model, which can provide qualitative description for surfaces with random roughness.

**Introduction**

Plasmon drag effect (PLDE) is a giant enhancement of photoinduced electric currents in metal films and nanostructures [1-9], associated with excitation of surface plasmon polaritons (SPPs). PLDE presents interest for various applications as it may provide the direct link for incorporation of plasmonic elements in electronic and optoelectronic devices. From a fundamental point of view, it can be considered in frames of plasmon-assisted momentum transfer in light-matter interaction in metal [2]. In first PLDE experiments [1,3], strong enhancement of photoinduced currents were observed in flat gold and silver films under conditions of surface plasmon polariton resonance excited in Kretschmann geometry [10]. Significant photo induced electric effects were reported in rough and nanostructured surfaces under direct illumination as well, with polarity and magnitude of electric signals dependent on angle of incidence, wavelength of illumination, light polarization, and nanoscale surface geometry [6-9].

Theoretical description [2,11] of PLDE is based on the electromagnetic momentum loss approach. The electromotive force (emf) in a continuous metal nanostructured surface can be found as an effect of the plasmonic pressure force acting on an electron, which can be written in components [11],

$$f_\alpha = \frac{1}{2} Re \left\{ \chi \left( E_x \partial_\alpha E_x^* + E_y \partial_\alpha E_y^* + E_z \partial_\alpha E_z^* \right) \right\} , \qquad (1)$$

where $\alpha = x, y, z$, $E_{x,y,z}$ are the electrical components of the optical field, and $\chi$ is the complex electric susceptibility of metal.

Recently it was shown [11] that this approach modified to take into account the electron thermalization time, $\tau_{therm}$, adequately describes strong electric currents induced under SPP resonance conditions in flat films. For a single-mode plasmonic field, the momentum transfer is directly related to the energy transfer, with the effective "plasmonic pressure" force acting on an electron as

$$F = \hbar k_{SPP} \frac{\tau_{therm}}{\tau} \frac{1}{n_e} \frac{Q_{abs}}{\hbar \omega} , \qquad (2)$$

where $\frac{Q_{abs}}{\hbar\omega}$ is the rate of plasmonic quanta absorption per unit volume, $k_{SPP}$ is the SPP wave vector, $\tau$ is the Drude collision time, and $n_e$ is the electron density. This approach can also be extended to surfaces with modulated profiles. This is important as it can provide a proper description of photoinduced electric effects in plasmonic nanostructures and guide the way to engineering plasmon induced electric responses with surface geometry.

It was theoretically shown [11] that under certain assumptions (a relatively small amplitude of surface modulation height, and laminar electric current), the electromotive force (emf) per unit length induced in a plasmonic structure can be presented as a sum of contributions from each plasmonic mode, with the contribution proportional to the absorbed power $Q_m$ and k-vector of each mode, $k_m$, as

$$U = \sum_m U_m = \frac{\tau_{therm}}{\tau} \frac{1}{n_e e} \sum_m \frac{\hbar k_m}{\hbar\omega} Q_m . \quad (3)$$

In order to prove validity of the theoretical approach [11] and explore its limitations, we use both the exact numerical approach (Eq. (1)) and simplified approach (Eq. (3)) for the description of PLDE in a metal grating-like structure, and compare theoretical predictions with experiment.

**Theoretical simulations**

The structure under consideration is a thin gold film of 60 nm thickness with a sine-wave profile, a periodicity $d$ = 538 nm and a modulation depth $2h$, which was varied in theoretical simulations. The sample schematic is shown in Figs. 1(a) and (b), and the optical setup is shown in Fig. 1(c).

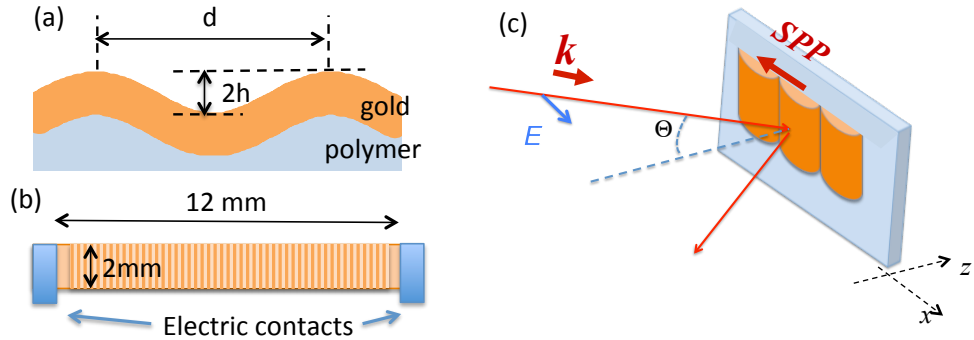

**Figure 1**. (a) Profile of the structure; (b) Geometry of the sample in PLDE experiments; (c) Optical setup. The film is illuminated with the p polarization.

The electromagnetic fields are calculated based on the Chandezon's method [12]. As expected, simulations predict surface plasmon resonance (SPR) [13] at the incidence angle, $\theta$,

$$k_{spp} = nG + k_x , \quad (4)$$

where $k_x = k \sin\theta$ is the projection of the optical $k$-vector on the $x$ axis, $G$ is the grating vector with a magnitude of $G=2\pi/d$, $n$ is an integer. The SPP $k$ vector closely corresponds to the estimations for flat films [13],

$$\frac{k_{SPP}}{k} = \xi = \sqrt{\frac{\varepsilon_m \varepsilon_d}{\varepsilon_m + \varepsilon_d}}, \qquad (5)$$

where $\varepsilon_m$ and $\varepsilon_d$ are correspondingly dielectric permittivities for metal and air.

Reflectivity of the film calculated for the whole optical range is shown in Fig. 2 (a) as a function of incidence angle, $\theta$, and light frequency for various amplitudes of the surface modulation.

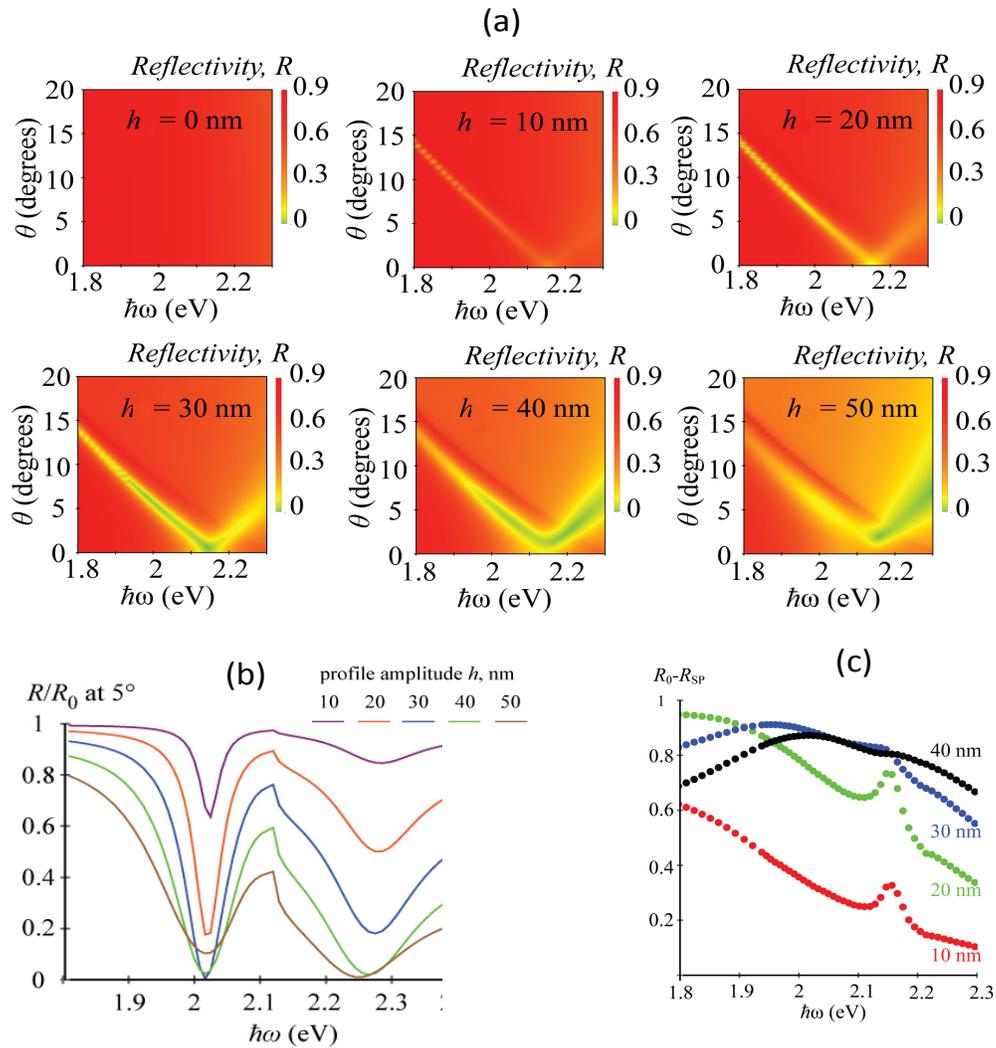

**Figure 2.** (a) Reflectivity as the function of the incidence angle and photon energy at various amplitudes of surface modulation; (b) Spectral dependence of $R/R_0$ at $\theta = 5$ deg; (c) Power absorbed at the resonance, $P_{SPP}$, at various $h$ as indicated.

Two SP branches are seen, which can be fitted with Eq. (4) with $n = -1$ or $+1$, corresponding to the plasmon propagating backward (against $k_x$, low frequency branch), and forward (in the direction of $k_x$, high frequency branch). These modes join each other at zero incidence angle and a frequency of ~ 2.15 eV, forming a standing plasmon wave with $k_{spp} = G/\xi$.

In Fig 2(b), the reflectivity $R$, normalized to the reflectivity of the flat film, $R_0$, is shown as a function of frequency for different modulation amplitudes at $\theta = 5$ deg. With an increase in the modulation amplitude, the resonance dips first become deeper and then broader with the further increase of $h$. Fig. 2(c) shows the spectral dependences of losses associated with the SPP excitation, $P_{SPP}$, which were calculated as the difference between the reflected intensity at the conditions of the SPP resonance for a particular $h$ and that of flat gold film ($h = 0$) at the same conditions. At low $h$ (10 nm and 20 nm), $P_{SPP}$ decreases with the increase in the frequency demonstrating a small peak at ~ 2.15 eV, Fig. 2(c). $P_{SPP}$ is the highest at $h \approx 30$ nm for most of the spectrum, indicating the most efficient SPP excitation. With the further increase in $h$ the spectral dependence $P_{SPP}(\omega)$ tends to, instead, increase with the increase of $\omega$.

The local distribution of the plasmonic pressure force (Eq. (1)) for the backward and forward SPPs is shown in Fig. 3(a). Magnitudes and directions of forces vary within the film, however the predominant direction corresponds to the SPP propagation direction for both the forward and backward SPP branches.

The photoinduced emf is calculated by the averaging of the pressure force over the volume of the metal film [11] as

$$\frac{U}{I} = \frac{\tau_{therm}}{\tau} \frac{L/I}{en_e} \frac{1}{h} \int_h \bar{f}_{Lx}(z) dz, \qquad (6)$$

where $I$ is the incident light intensity, $L$ is the diameter of the illuminated spot, and $\tau_{therm}$ is the electron thermalization time [14,15].

Pure photonic drag [16], $U_0$, is relatively small and plays a role only at the blue part of the spectrum as will be shown below. The plasmonic emf, $U - U_0$, calculated at $\tau_{therm} = 270$ fs and $L = 2$ mm is plotted in Fig. 3 (b). The polarities of $U - U_0$ at red and blue parts of the spectrum are opposite to each other, in each case corresponding to the drift of electrons in the direction of SPP propagation. The widths and magnitudes of the emf peaks change with the variation of $h$ in similar way to the dips in the reflectivity (Fig 2 (b)).

For further analysis, the ratio of the peak emf at SPR conditions and maximum losses associated with SPR are plotted in Fig. 3(c) for the whole optical range. The curves are almost flat, corresponding to $C = \frac{1}{I} \frac{U-U_0}{R_0-R} = 2.2$ -2.5 mV/(MW/cm$^2$) at small modulation amplitudes, increasing to ~ 3-3.5 with an increase in $h$. They are negative at the red part of the spectrum, and positive at blue, corresponding to backward or forward plasmon drag. The polarity switches at ~ 2.15 eV where SPPs are excited in both directions with equal efficiency, providing net zero momentum to electrons. Near the switching point the curves with high $h$ demonstrate a small peak.

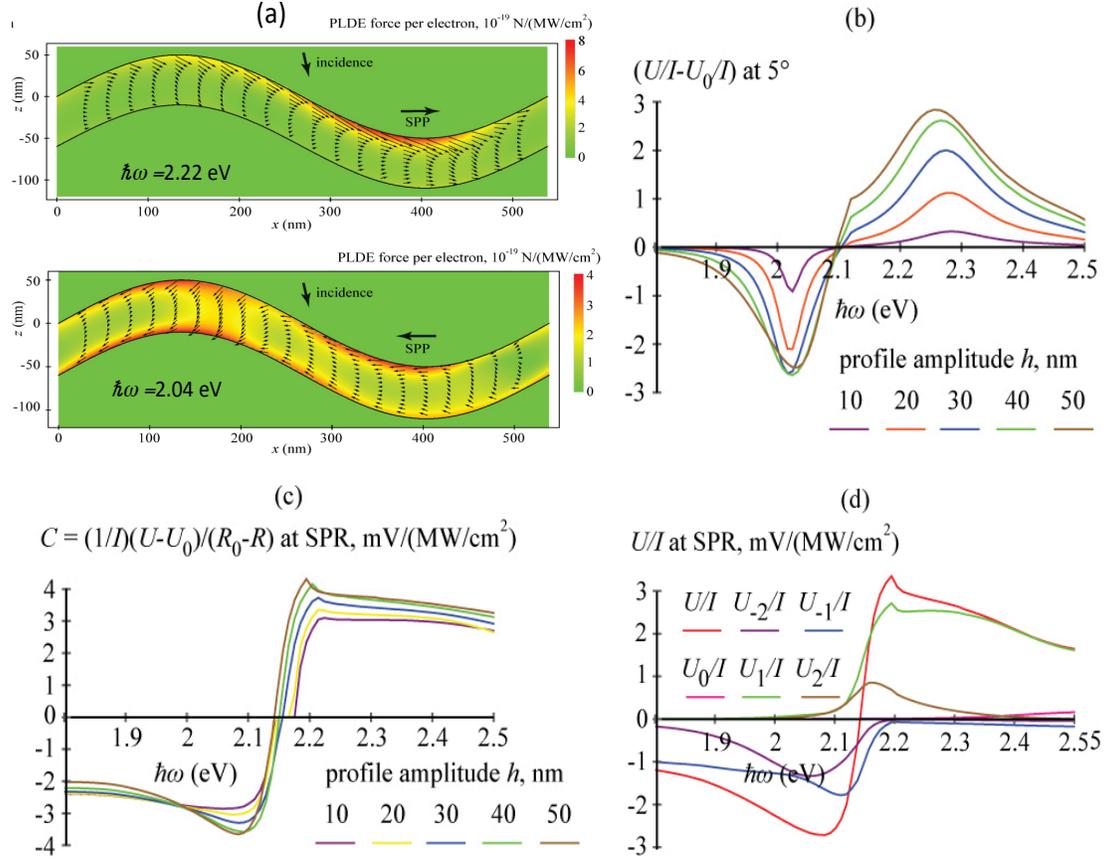

**Figure 3.** Local distributions of the plasmonic pressure force at the resonance conditions for forward and backward SPPs; (b) emf at $\theta= 5$ deg.; (c) Ratio $C=\frac{1}{I}\frac{U-U_0}{R_0-R}$ at SPR for various modulation amplitudes; (d) Different contributions to the emf at $h=50$ nm.

The flatness of the curves indicates that the momentum transfer from plasmons to electrons is directly proportional to the energy transfer, which correlates well with the conclusions of the theory [11] predicting a direct proportionality between the plasmon induced emf and absorption at relatively small modulations of the metal surfaces ($h << d$) and laminar electron flow. With an increase in the modulation amplitude, the ratio, C, becomes slightly higher, which can be partially explained with the contributions of higher orders of diffraction, Fig. 3(d). Note the effect of the photonic pressure, $U_0$ is significantly smaller than plasmonic pressure contributions, and is noticeable at high frequencies only, Fig. 3(d).

**Experiment**
Let us now compare the theory with the experiment. The experimental structure is strip shaped (Fig. 1(b)) with a width of 2 mm and a length of 12 mm. It was produced with thermal deposition of gold through a mask onto a polymer grating, which was fabricated with the holographic lithography method [17,18]. The gold film had a thickness of 60 nm ± 2 nm, the period of the modulation of 538 nm, and the modulation depth varied from 40 nm to 60 nm. The orientation of the grating grooves was perpendicular to the long side of the strip. Two electric contacts were attached on the opposite ends of the sample. The reflectivity of the sample was measured in the setup shown in Fig. 1(c) under illumination with the He-Ne laser at 632.8 nm. The plasmon drag effect was studied in the same optical setup under pulsed laser light illumination using the OPO laser light with an incident power of ~ 0.3 mJ per pulse and ~ 5 ns pulse duration. The diameter of the laser spot was of about 2.5 mm, which was slightly larger than the width of the strip. The photoinduced electric signal was recorded with Tektronix oscilloscope with 50 Ω resistance.

In Fig. 4(a), a dip in the reflectivity is observed at ~ 8 degree incidence angle indicating the SPR conditions. The photoinduced voltage is shown as a function of $\theta$ at the wavelengths of illumination $\lambda$ = 630 nm (Fig. 4 (b)) and 550 nm (Fig 4(c)). In our plot and discussion below, the positive signal at positive angles, and negative signal at negative angles correspond to the direction of electron drift in the direction of $k_x$. At $\lambda$ = 550 nm, the polarity of the signal corresponds to the drift in the direction of $k_x$ at the whole range of incidence angles. At $\lambda$ = 630 nm, the signal has the opposite polarity at small angles (electrons drift against $k_x$). However, at large angles the polarity of the signal switches back to the polarity corresponding to the electron drift in the direction of $k_x$.

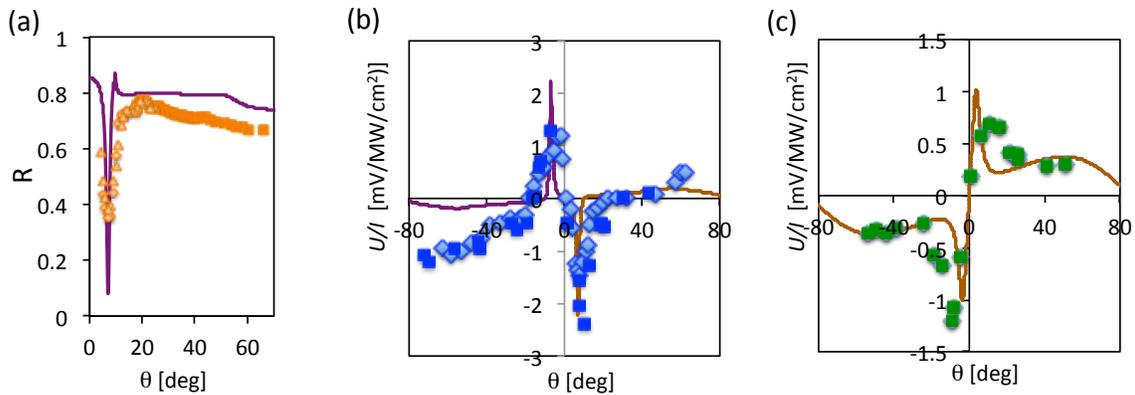

**Figure 4.** (a) Reflected intensity at 632.8 nm; Photoinduced electric signal at (b) $\lambda$ = 630nm and (c) 550nm. Points are experimental data, different symbols correspond to different experimental runs. Solid traces are numerical simulations.

The peaks in both graphs correspond to the SPPs propagating forward at illumination at 550 nm ($k_{spp} = G+k_x$) and backward at illumination at 630 nm ($-k_{spp} = k_x - G$). The magnitude of the electric signal at low angles can be fit with the numerical simulations assuming $h$ = 20 nm, see solid traces in Fig. 4. However, the experimental curves are significantly broader than the theory, which can be ascribed to effects of roughness. An additional feature clearly seen in Fig. 4 (b) is the monotonous growth of the signal magnitude with the increase in $\theta$, with the polarity of the signal corresponding to the drift of the electrons in the direction of $k_x$. Such a behavior is similar to what was observed in rough metal films [6, 9], and may be associated with SPPs excited due to additional roughness.

## Discussion: Equivalent circuit model

While accounting for random roughness presents a challenge for exact numerical simulations, let us now make an attempt to discuss the photoinduced emf in frames of the equivalent circuit model, and test the possibility of explaining the signal observed at large angles with SPPs excited due to roughness. The goal of the consideration below is not to achieve full correlation with numerical calculations and experimental data, but rather to interpret the results and explore general tendencies.

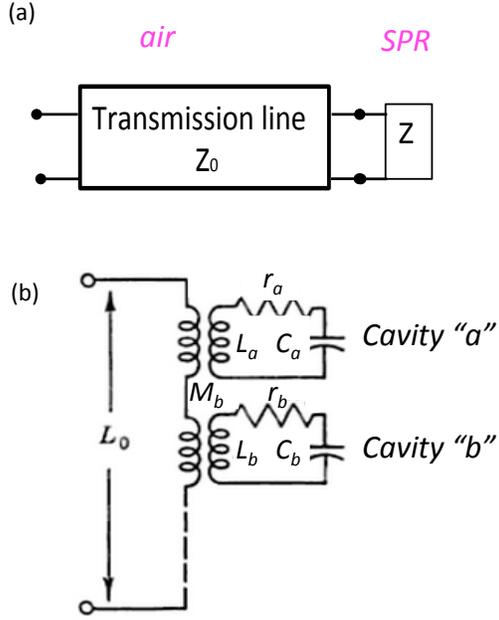

**Figure 5.** (a) Equivalent circuit schematic; (b) Coupling of the resonant cavities.

Consider a circuit consisting of a transmission line coupled through the loop with the resonance circuit consisting of two cavities having different resonance frequencies, see Fig 5. The reflectivity from such a resonant circuit can be found as

$$|\Gamma|^2 = \left|\frac{Z_0 - Z}{Z_0 + Z}\right|^2, \quad (7)$$

where $\Gamma$ is the complex reflection coefficient, $Z_0$ is the impedance of the transmission line, and $Z$ is the equivalent input impedance [19],

$$Z = i\omega L_0 + \frac{i\kappa_a \omega^2}{\omega_a^2 - \omega^2 + \frac{i\omega\omega_a}{Q_a}} + \frac{i\kappa_b \omega^2}{\omega_b^2 - \omega^2 + \frac{i\omega\omega_b}{Q_b}}, \quad (8)$$

$$\kappa_{a,b} = \frac{\omega \, M_{a,b}^2}{L_{a,b}}. \quad (9)$$

Here $M_{a,b}$, $L_0$ and $L_{a,b}$ are the mutual inductance and equivalent inductances of the line and the resonant circuits respectively, $r_{a,b}$ and $C_{a,b}$ are the equivalent resistance and capacitance of the resonant circuits, $\omega_{a,b} \approx (L_{a,b} C_{a,b})^{-1/2}$, and $Q_{a,b} = \omega_{a,b} L_{a,b}/r_{a,b}$. Energy exchange between incident light and the SPPs in optical experiments can be viewed in terms of the energy exchange between the transmission line and two resonant cavities corresponding to the two branches of SPR: the cavity "a" describes the excitation of the forward SPP, and the cavity "b" describes the "backward" SPP. Coupling is provided with the sine-wave profile of the film. We assume $\kappa_a/r_0 = \kappa_b/r_0 \propto h^2$, which, in terms of an equivalent circuit, corresponds to the linear dependence of the mutual inductance on the modulation amplitude, $M_{a,b} \propto h$. As shown below, this assumption well corresponds to the results of the numerical calculations. Let us also take into account the fact that for the same incidence angle, the dip in reflectivity is broader for higher photon energies, Fig.

2(b), and assume that $Q$-factor of the cavities quickly decreases with the increase in the resonance frequency as $Q(E) = B/\omega^6$, where $B$ is the constant.

Equations (7)-(9) predict dips in the reflectivity at the resonance frequencies, Fig. 6(a). The curves are obtained assuming a predominately active resistance for the transmission line, $Z_0 = r_0 \gg \omega L_0$, the resonance frequencies 2.015 eV and 2.285 eV, the $Q$-factors of 48 and 22.5, and various $\kappa/r_0$, proportional to $h^2$, where $h$ varies from 10 nm to 50 nm. As one can see, the equivalent circuit model predicts the results very similar to numerical simulations, including the initial increase in depth with the increase of $h$ from 10 nm to 30 nm, and broadening of the dip with a further increase of $h$. Variation of the incidence angle $\theta$ can be approximately described with the "tuning" of the resonance frequencies of cavities. Varying both resonance frequencies toward each other with the point of coincidence at 2.15 eV, the spectral dependences of 1-$|\Gamma|^2$ were calculated, Fig. 6(b). In similarity with Fig. 2(c), the peak at ~ 2.15 eV is seen for cavities with low coupling. At high $h$, the spectral dependence flips, which also corresponds to the numerical results presented in Fig. 2c.

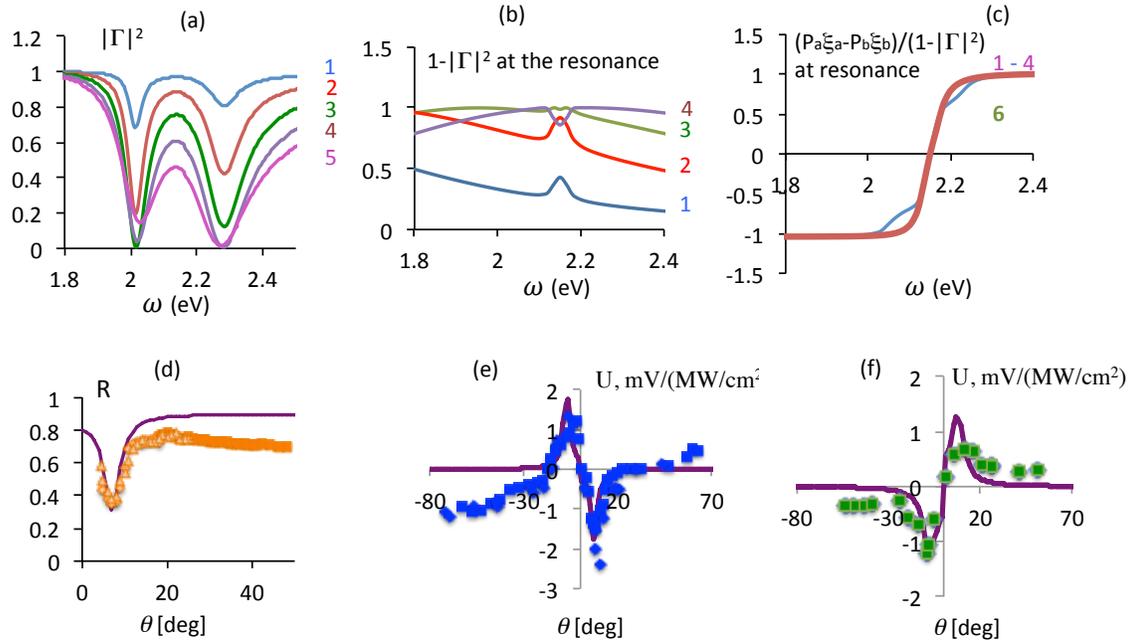

**Figure 6**. (a) Reflection from the cavity with resonances at $\omega_a$, = 2.015 eV, $\omega_b$,=2.285 eV, $B$=3200 eV$^6$; (b) Frequency dependence of the resonant dip depth; (c) Ratio of the emf and absorbed power at the resonance. Traces correspond $\kappa/r_0$ = 0.001 (Trace 1), 0.004 (2), 0.009 (3), 0.016 (4), 0.025 (5), and 0.036 (Trace 6). Fitting the experimental angular dependences of reflectivity and the emf with the equivalent circuit model at (d, e) $\lambda$= 630 nm, (f) $\lambda$= 550 nm. $B$= 1000 eV$^6$, $\kappa/r_0$ = 0.08, $\alpha$ = 2.2 mV/(MWcm$^2$) /P, $R$ = 0.9$|\Gamma|^2$, $\xi$ = 1.05.

Let us now find the emf which is expected to be proportional to the power absorbed at the resonance cavity, $P_{a,b}$, and has the polarity determined by the direction of the SPP propagation. In the equivalent circuit with two cavities corresponding to SPPs excited in backward and forward directions, the emf is a sum of contributions from the cavities with the opposite sign as

$$U = \alpha \, (P_a \xi_a - P_b \xi_b) = \alpha \, |J|^2 \, (Re(Z_a) \, \xi_a - Re(Z_b) \, \xi_b). \tag{10}$$

where $J$ is the current, $\alpha$ is the proportionality constant, $\xi_a$ and $\xi_b$ are parameters describing the ratio $k_{spp}/k$ at the resonance frequencies $\omega_a$ and $\omega_b$ correspondingly.

Assuming weak dependence of $\xi(\omega)$, the curves $U/(1-|\Gamma|^2)$ shown in Fig 6(c) are flat at low $\kappa/r_0$ in similarity to the results of numerical simulations, Fig. 3(c). Some variation in shape is seen near the switching point for cavities with high $\kappa/r_0$; however the character of this variation is different from what is predicted by numerical simulations above, and can be simply explained with the overlapping of significantly broadened resonances near the switching point, providing opposite contributions to the emf. Thus, the equivalent circuit model reasonably well describes the predictions of strict numerical simulations for the reflectivity and emf at relatively small modulation amplitudes ($h<30$ nm), and departs from that at large $h$. In Fig 6 (d-f), this model is able to fit the experimental reflectivity and photoinduced electric signal at the SPR conditions with a proper choice of the equivalent circuit parameters.

Now let us take into account plasmons excited by the roughness in terms of additional resonance cavities connected in series. In order to see qualitative behavior, we make several simplifying assumptions and consider only the SPPs with $n = \pm 1$. Presenting the surface profile as a sum of the spatial harmonics, $\sum F(G_i)$, each grating vector, $G_i$ can be associated with excitation of forward and backward SPPs with efficiency dependent on the corresponding amplitude of the spatial harmonics, $F(G_i)$. In similarity with the previous fitting (Fig. 6), $\kappa \propto F(G_i)^2$. At a particular angle $\theta$, only cavities with the resonance frequency close to the illumination frequency, $\omega = \omega_0 \pm \Delta\omega$, absorb the energy. Assuming $\xi \approx const$, the grating vectors of such cavities can be found as

$$G = \frac{\omega_0(\xi \pm sin\theta)}{c}. \tag{11}$$

Losses in such cavities contribute to emf positively at $G < \omega_0\xi/c$, and negatively at $G > \omega_0\xi/c$. The total emf associated with roughness is determined by

$$U_r \propto \left( F^2\left(\frac{\omega_0(\xi - sin\theta)}{c}\right) - F^2\left(\frac{\omega_0(\xi + sin\theta)}{c}\right) \right) \xi. \tag{12}$$

From Eq. (12) the polarity and magnitude of the roughness-associated contribution to emf is determined by the spectrum of spatial harmonics. If higher harmonics prevail, the polarity of the signal would correspond to the drift of electrons against $k_x$, as was observed in nominally flat films in the experiments with prisms [3].

Let us consider a random roughness with the Gaussian distribution,

$$F(G) = f_o \, \delta^{-1} exp\left(-\frac{G^2}{\delta^2}\right), \tag{13}$$

where $f_o$ is a constant, and $\delta$ is the half-width of the distribution, Fig 7(a). Substituting Eq. (13) to Eq (12),

$$U_r \propto \frac{f_o^2}{2\delta^2} \exp\left[-\frac{2\omega_0^2}{c^2\delta^2}(\xi^2 + sin^2\theta)\right] \sinh\left[\frac{4\omega_0^2\xi\ sin\theta}{c^2\delta^2}\right]\xi, \qquad (14)$$

while the power losses in such cavities are

$$P_r \propto \frac{f_o^2}{2\delta^2} \exp\left[-\frac{2\omega_0^2}{c^2\delta^2}(\xi^2 + sin^2\theta)\right] \cosh\left[\frac{4\omega_0^2\xi\ sin\theta}{c^2\delta^2}\right]. \qquad (15)$$

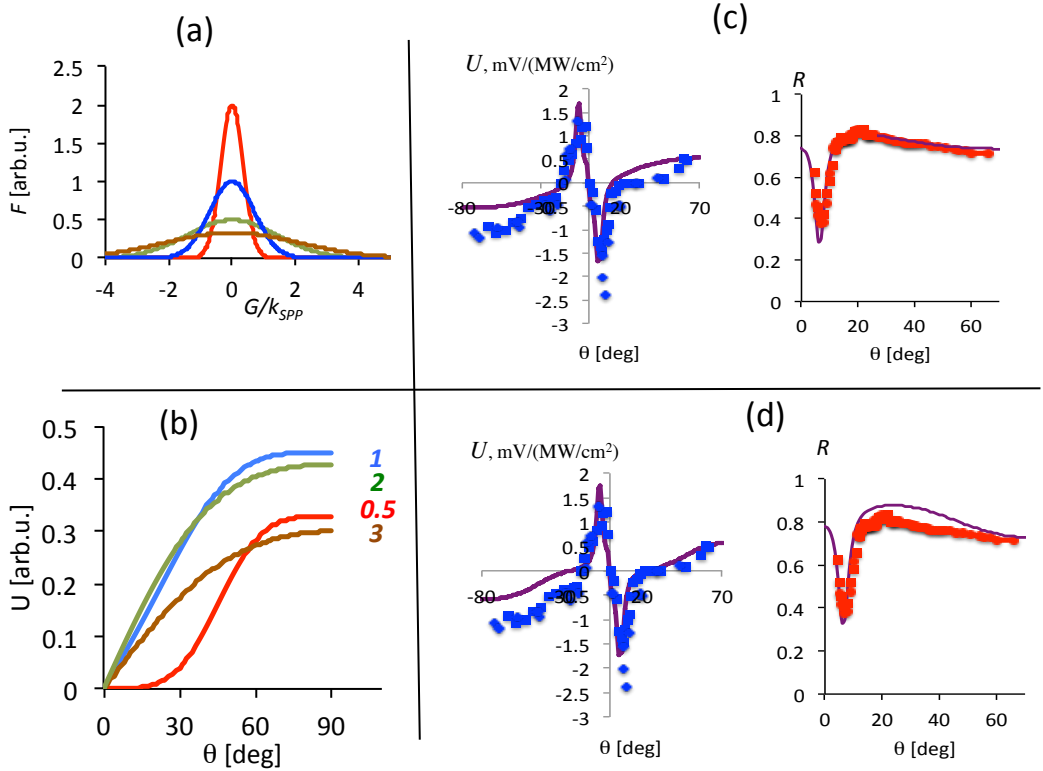

**Figure 7**. (a) Spectrum of spatial harmonics and (b) contribution to the emf (bottom) at $\delta/k_{spp}$ = 0.5, 1, 2, and 3 as indicated. (c, d) Fitting experimental $R$ and $U$ with the equivalent circuit model and account for roughness at (c) $\delta/k_{spp}$ =1 and (d) $\delta/k_{spp}$ =0.5.

In figure 7 (b), the photoinduced emf associated with roughness is plotted for the various $\delta$. At a relatively broad distribution, $\delta > k_{sp}$, the angular dependence is similar to the experimental dependences observed in the rough films [6], where the dependence $U \propto sin(\theta)$ is reported. In Figs. 7(c, d) we make an attempt to fit our experiment at 630/632 nm, making an assumption that the total change in reflectivity at high angles is due to additional SPPs excited in $x$ or $-x$ direction with the $k$-vector being the same as in flat films, $\xi \sim 1$. As one can see, the account for a roughness with the relatively narrow distribution, $\delta/k_{spp}$ = 0.5 : 1 can reasonably well fit both the angular dependences of $R$ and $U$ in our sample. The possible source of such a distribution can be associated with the variation of the modulation height in the experimental sample. However, the quantitative fitting of the data for large incidence angles is still problematic even for such strong assumptions. One of the possible reasons of the discrepancy of the experiment and this particular fitting is the participation of plasmonic modes with high *k*-vectors excited at corrugated surfaces [20], which may have correspondingly higher contributions to the photoinduced voltage, $U_i \sim \xi_i$. Note that localized surface plasmons (LSP) commonly associated with roughness, are non-propagating and do not contribute to PLDE. Another factor is the limitations of the simplified consideration for surfaces with relatively high slopes of the surface modulation. As seen from Figs. 3(c) and 6(c), the predictions of the exact numerical simulations and equivalent circuit model for high modulation amplitudes have different behavior around the switching point.

In conclusion, PLDE in a thin gold film with a sine-wave profile is analyzed theoretically and studied experimentally. Both the numerical simulations based on modified electromagnetic momentum loss approach and the equivalent circuit model based on the simplified approach reasonably well describe the emf associated with propagating SPPs at small amplitudes of surface modulations. Effect of roughness is considered in terms of additional plasmonic modes excited at rough surfaces. Limitations of the simplified consideration are discussed.

**Acknowledgments**

The work was partially supported by PREM grant no. DMR 1205457 and the Army Research Office (ARO) grant W911NF-14-1-0639. M. D. was supported by funds from the Office of the Vice President for Research & Economic Development at Georgia Southern University.